# Charge Storage in Cation Incorporated α-MnO$_2$

Matthias J. Young[1,‡], Aaron M. Holder[1,2,‡], Steven M. George[2,3], and Charles B. Musgrave[1,2,*],

[1]Department of Chemical and Biological Engineering, [2]Department of Chemistry and Biochemistry, [3]Department of Mechanical Engineering, University of Colorado, Boulder, Colorado 80309

**ABSTRACT:** Electrochemical supercapacitors utilizing α-MnO$_2$ offer the possibility of both high power density and high energy density. Unfortunately, the mechanism of electrochemical charge storage in α-MnO$_2$ and the effect of operating conditions on the charge storage mechanism are generally not well understood. Here, we present the first detailed charge storage mechanism of α-MnO$_2$ and explain the capacity differences between α- and β-MnO$_2$ using a combined theoretical electrochemical and band structure analysis. We identify the importance of the band gap, work function, the point of zero charge, and the tunnel sizes of the electrode material, as well as the pH and stability window of the electrolyte in determining the viability of a given electrode material. The high capacity of α-MnO$_2$ results from cation induced charge-switching states in the band gap that overlap with the scanned potential allowed by the electrolyte. The charge-switching states originate from interstitial and substitutional cations (H$^+$, Li$^+$, Na$^+$, and K$^+$) incorporated into the material. Interstitial cations are found to induce charge-switching states by stabilizing Mn-O antibonding orbitals from the conduction band. Substitutional cations interact with O[2p] dangling bonds that are destabilized from the valence band by Mn vacancies to induce charge-switching states. We calculate the equilibrium electrochemical potentials at which these states are reduced and predict the effect of the electrochemical operating conditions on their contribution to charge storage. The mechanism and theoretical approach we report is general and can be used to computationally screen new materials for improved charge storage via ion incorporation.

## Introduction

The promise of simultaneous high energy and power density has sparked growing interest in electrochemical supercapacitors.[1,2] In addition to double-layer capacitance,[3] electrochemical supercapacitor materials are known to store charge at the surface and near-surface region through electrochemical charge transfer processes. Unlike bulk phase-changes, the processes leading to charge storage in these materials occur at extraordinary rates with cycle lifetimes of >10$^5$ cycles.[4,5] These charge transfer processes have been described as occurring at broadened equilibrium potentials that overlap to result in a nearly linear dependence of charge transferred (Q) versus applied potential (Φ), and thus a nearly constant capacitance ($C = \frac{dQ}{d\Phi}$).[6] This electrochemical behavior mimics the rectangular-shaped cyclic voltammetry (CV) curves of double-layer or parallel-plate capacitors and is therefore termed "pseudocapacitance". Because electrochemical supercapacitor electrodes based on manganese dioxide (α-MnO$_2$) exhibit a high experimental specific capacity and are composed of low toxicity earth-abundant elements, they have become an attractive choice for commercialization.[7–9]

### Model Description

In this study, we reveal the fundamental basis for fast and highly reversible charge storage in α-MnO$_2$. Our analysis is performed within a widely transferrable band diagram framework to evaluate the electrochemical mechanism of charge storage, as illustrated in Figure 1. In this framework, incorporated cations are modeled as extrinsic impurities using solid-state defect theory to identify the cation-induced electronic levels within the band gap (charge-switching states). These cation induced electronic gap states store charge by accepting and donating electrons as the applied potential shifts the Fermi level ($\varepsilon_f$) above and below their charge switching potentials, respectively. The role of charge-switching by interstitial cations and other defect states has been studied for a variety of applications,[10–13] however the approach we apply here to identify thermodynamically accessible charge-switching states within an electrochemical framework has not been previously used to study charge storage materials which incorporate cations. Our approach evaluates the alignment of the potential at which charge-switching states accept or donate electrons with the scanned potential window (SPW) allowed by the electrolyte, beyond which the electrolyte undergoes oxidation or reduction.

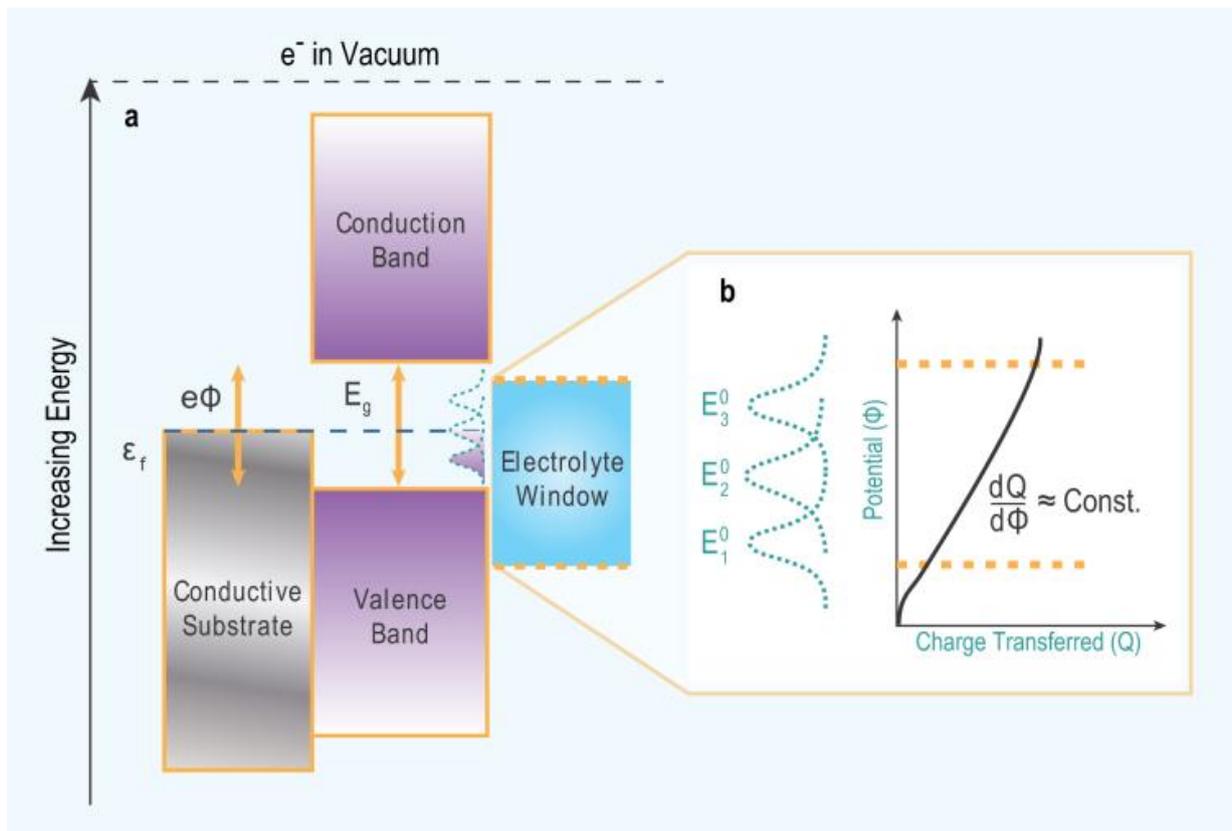

**Figure 1. Band diagram and charge-switching alignment of an electrochemical supercapacitor electrode.** (**a**) Simplified band diagram description of the electrode components. The electrode is composed of an electrically conducting current collector, an electrochemically active electrode material with a bulk band gap ($E_g$) and an electrolyte with a known stability window, outside of which it undergoes either oxidation or reduction. Electrochemically active electrodes possess defects with electronic energy levels within the band gap that undergo electrochemical reduction/oxidation to store charge. Consequently, we define these defect levels that lead to charge storage as "charge-switching states". The operating window is limited to the region of potentials where the electrode band gap and electrolyte stability window overlap. (**b**) Charge-switching states at potentials ($E_n^0$) are broadened by interactions with their surroundings resulting in a nearly linear relationship between potential and stored charge, and a constant capacity.

The equilibrium potentials ($E^o$) and thermodynamics of these charge-switching states are computed using quantum mechanical calculations employing periodic density functional theory with a screened non-local exchange-correlation functional (HSE06)[14,15]; computational details are provided in the Supporting Information (SI). We utilized a bulk description to calculate cation induced gap (defect) states and incorporated the influence of the electrolyte via pH and cation chemical potential corrections to simulate the environment near the electrode surface. This framework identifies the importance of the band gap, work function, and point of zero charge of the electrode material, as well as the pH and stability window of the electrolyte as the primary properties that determine the viability of a given electrode material for charge storage.

### α-MnO₂ as a Charge Storage Material

Until now, no detailed mechanism has been developed for the fast, reversible charge storage observed in α-MnO₂. However, many experimental observations pertaining to the mechanism have been documented and any proposed mechanism must be reconciled with these observations to be viable. *Ex situ* X-ray photoelectron spectroscopy (XPS) studies indicate that the charge storage process in aqueous electrolyte involves the formation and disappearance of hydroxyl groups on the electrode surface and a concomitant change in the formal oxidation state of Mn from $3^+$ to $4^+$.[8,16] Also, an observed pH dependence of charge storage suggests that processes mediated by protons account for ~1/2 of the specific capacity of α-MnO₂.[17,18] Other cations, in addition to protons, are also involved in charge storage and are thought to contribute the remaining capacity.[8,16,17,19] Both protons and larger cations have been described to intercalate and deintercalate[19,20] through bulk α-MnO₂ during charging and discharging.[21] These observed sources of charge storage have been combined to express the reactions involved in the charge storage mechanism of α-MnO₂ as[8,16,17,19]

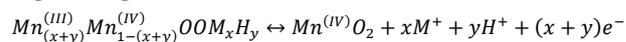

$$Mn_{(x+y)}^{(III)} Mn_{1-(x+y)}^{(IV)} OOM_xH_y \leftrightarrow Mn^{(IV)}O_2 + xM^+ + yH^+ + (x+y)e^-$$

where $M^+$ represents a singly charged cationic species. A similar expression describes reactions between α-MnO₂ and multiply charged cations involved in charge storage.

The specific capacity of MnO$_2$ strongly depends on its film thickness and crystalline phase, with specific capacitances reported for thin-film α-MnO$_2$ of ~1000 F/g and as high as 1380 F/g, which corresponds to 1.1 electrons per Mn center.[16,22] In contrast, crystalline α-MnO$_2$ materials exhibit a bulk capacitance of only ~200 F/g, while the β-MnO$_2$ crystalline phase has a meager bulk capacitance of ~10 F/g.[23,24] One explanation given in the literature for the capacity differences between α-MnO$_2$ and β-MnO$_2$ is the tunnel (also called channel) sizes of their crystal structure; α-MnO$_2$ has larger tunnel sizes that have been suggested to enhance ion diffusion (e.g. proton conductivity) and provide additional adsorption sites to accept cations.[16,25,26] However, the pervasiveness of surface and bulk protons incorporated into metal oxides without tunnel structures[13,27,28] and the observed charge storage in crystal materials without tunnel structures, such as RuO$_2$,[8] suggests that the smaller tunnel size alone does not preclude β-MnO$_2$ from storing charge by an analogous mechanism. Our work identifies dramatic differences in the electronic properties of α-MnO$_2$ and β-MnO$_2$ and elucidates a detailed charge storage mechanism that explains the capacity difference between these crystalline phases, as well as the higher capacity of thin-film α-MnO$_2$.

## Results and Discussion

### Band Alignments of α-MnO$_2$ and β-MnO$_2$

The band alignments for α- and β-MnO$_2$ are plotted on an absolute scale with respect to an electron in vacuum in Figure 2. The alignments were determined using calculated pH-corrected work functions of 7.7 and 8.1 eV for the (110) surfaces and a calculated indirect band gap of 2.7 eV and a direct band gap of 1.6 eV for α- and β-MnO$_2$, respectively. SI Section A describes how these band gaps were determined by benchmarking HSE06 against GW$_0$ quasiparticle calculations. The SPW within the allowed aqueous electrolyte window is also displayed for comparison. This analysis shows that the experimental SPW (0 to 1 V vs. Ag/AgCl)[29,30] does not overlap with the band gap of β-MnO$_2$. For a β-MnO$_2$ electrode, the SPW is entirely in the conduction band, so the potential at a β-MnO$_2$ electrode will simply be pinned in the conduction band over the SPW and the charge-switching states in β-MnO$_2$ will not switch charge at experimentally accessible potentials. Thus, β-MnO$_2$ is not an effective charge storage material in aqueous electrolyte, as observed experimentally. In contrast, the band gap of α-MnO$_2$ overlaps with the SPW. Therefore, α-MnO$_2$ will be a viable charge storage material if its charge-switching states lie within the portion of its band gap that overlaps the SPW of the electrolyte, and its charge-switching states are thermodynamically favorable.

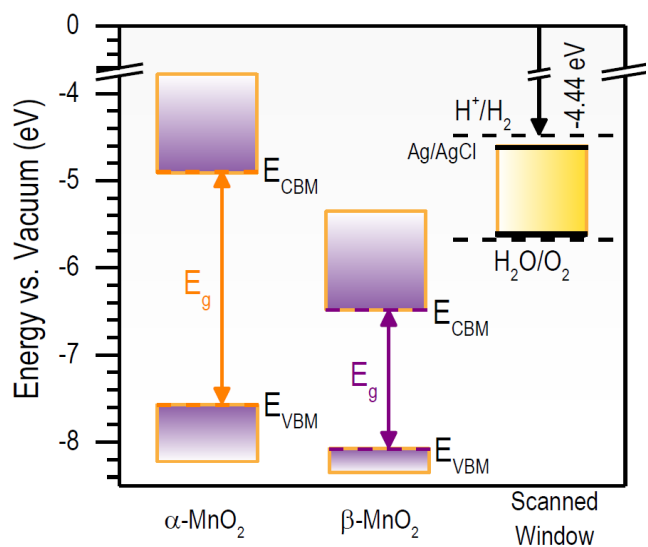

**Figure 2.** The absolute band edge energies of α-MnO$_2$ (left) and β-MnO$_2$ (center) and the electrochemical scanned potential window (SPW) for MnO$_2$ in aqueous electrolyte at pH = 7.4 (right).

### Cation Induced Electronic Charge-Switching States

The possible cation incorporation locations we investigate that may lead to induced electronic charge storage states in MnO$_2$ are depicted in Figure 3a. α-MnO$_2$ possesses both small 1×1 and larger 2×2 tunnels, whereas the β-MnO$_2$ structure only exhibits the smaller 1×1 tunnels (tunnel sizes are in units of MnO$_2$ octahedral blocks). The views along the [001] directions show the locations of interstitial incorporated cations ($^{2\times2}M_i$ and $^{1\times1}M_i$). The $^{2\times2}M_i$ cations are located in the larger 2×2 tunnels, whereas $^{1\times1}M_i$ cations are positioned in the smaller 1×1 tunnels. Also depicted are manganese vacancies ($V_{Mn}$) and their cationated form ($M_{Mn}$), which are Mn vacancies occupied by an M$^+$ cation. These incorporated cations (substitutional and interstitial) all potentially lead to electronic states that store charge at potentials within both the band gap and the SPW of the electrolyte. In addition, Figure 3a also illustrates oxygen vacancies ($V_O$) for reference. Oxygen vacancies prevalent in manganese oxides are interesting for other applications, but are not expected to contribute to cation incorporation charge storage. The strategy we use here for studying incorporated cations expands upon existing techniques for calculating equilibrium potentials in battery materials[31,32] by evaluating multiple possible charge states for a given cation incorporated into the host electrode material.

A comprehensive electrochemical thermodynamic picture for defects (oxygen vacancies and incorporated cations) in α-MnO$_2$ at a simulated pH of 7.4[23,33] is shown in the plot of formation energy versus potential in Figure 3b. In this plot, defects with a particular charge become thermodynamically favorable when their formation energies ($\Delta E_f$) are less than zero. The potentials marked in Figure 3b indicate the potentials at which it is thermodynamically favorable for defects to accept an additional electron and therefore store charge. By scanning the applied bias to more reducing potentials (from left to right), an electron is transferred to the state at each point marked on the line, provided that the defect is present. Because the potential will be pinned at the valence band and conduction band edges, the formation energies are not plotted beyond the band gap. The accuracy of the predicted charge-switching model is corroborated by multiple experimental observations; a) We predict that protons will occupy Mn vacancies, creating Ruetschi-type defects as observed experimentally;[34] b) Our results show Mn vacancies become favorable at $\Phi < 0.1$, and O vacancies become favorable at $\Phi > 0.8$, resulting in a predicted stable potential window similar to the experimental potential window of $0 < \Phi < 1.0$ V vs. a Ag/AgCl reference electrode[29,30]; and c) We show that interstitial protons ($^{1\times1}H_i$ and $^{2\times2}H_i$) and protonated Mn vacancies ($H_{Mn}$) undergo charge-switching at potentials within the SPW of the electrolyte, indicating that these defects lead to proton mediated charge storage in α-MnO$_2$.[16–18]

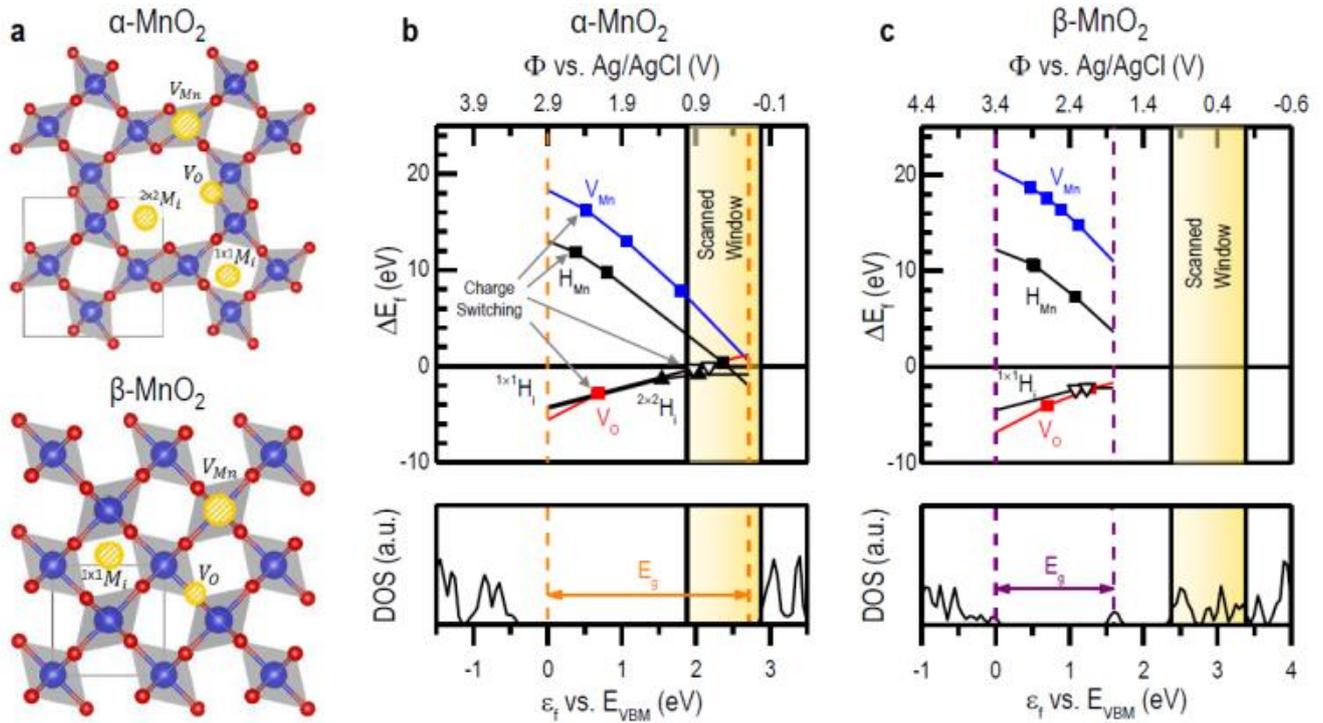

**Figure 3.** Vacancy and proton induced charge-switching states of α-MnO$_2$ and β-MnO$_2$. (**a**) Crystal structures and defect sites of α-MnO$_2$ and β-MnO$_2$ with manganese atoms shown in blue and oxygen atoms shown in red. Results for (**b**) α-MnO$_2$ and (**c**) β-MnO$_2$ showing formation energies ($\Delta E_f$) and location of charge-switching potentials as a function of applied bias ($\Phi$). The band gaps are indicated by the dashed vertical lines and the projected density of states (PDOS) for the negatively charged interstitial proton defects are shown in the bottom panel. (see Figure 4 for the nature of the charge-switching states).

The calculated density of states (DOS), band gap and the SPW allowed by the aqueous electrolyte are also displayed for reference for α-MnO$_2$ in Figure 3b. The indirect band gap of α-MnO$_2$ overlaps with the SPW and also contains charge-switching states within the SPW. Thus, α-MnO$_2$ possesses the requisite properties for charge storage. In contrast, our calculations show that the charge-switching states that lie within the band gap of β-MnO$_2$ are outside the SPW (see Figure 3c). This analysis extends beyond the simplistic analysis limited to examining the tunnel sizes of each phase and more accurately explains the charge storage properties of each material. These results demonstrate why α-MnO$_2$ is a viable charge storage material and why β-MnO$_2$ does not effectively store charge in aqueous electrolyte, as observed experimentally.

**Dominant Interstitial Cation Mechanism in α-MnO$_2$**

The dominant mechanism of charge storage in α-MnO$_2$ results from interstitial cations incorporated into the structure (illustrated in Figure 4a), as has been suggested in other work.[25,35] Our results show that interstitial protons or cations, M$^+$ localize near oxygen atoms. The Coulombic field of M$^+$ distorts the electronic distribution of the adjacent Mn-O bond, which stabilizes the corresponding Mn-O antibonding d-p $\pi^*$ orbitals from the conduction band into the band gap. The projected DOS (PDOS) (Figure 4b) shows that these Mn-O antibonding states are primarily composed of Mn[3d] orbitals, which agrees with XPS results that indicate a change in the Mn valency from 4$^+$ to 3$^+$ over the SPW.[16] Applying a potential bias of sufficient magnitude and direction to the electrode populates or depopulates these stabilized antibonding orbitals leading to charge-switching of these interstitial defects. For thin-film α-MnO$_2$ at a pH of 7.4, Li$^+$, Na$^+$ and K$^+$ have favorable interstitial formation energies for the $^{2\times2}M_i$ site, whereas formation of interstitial H$^+$ is favorable for both the $^{1\times1}M_i$ and $^{2\times2}M_i$ sites (Figure 4b).

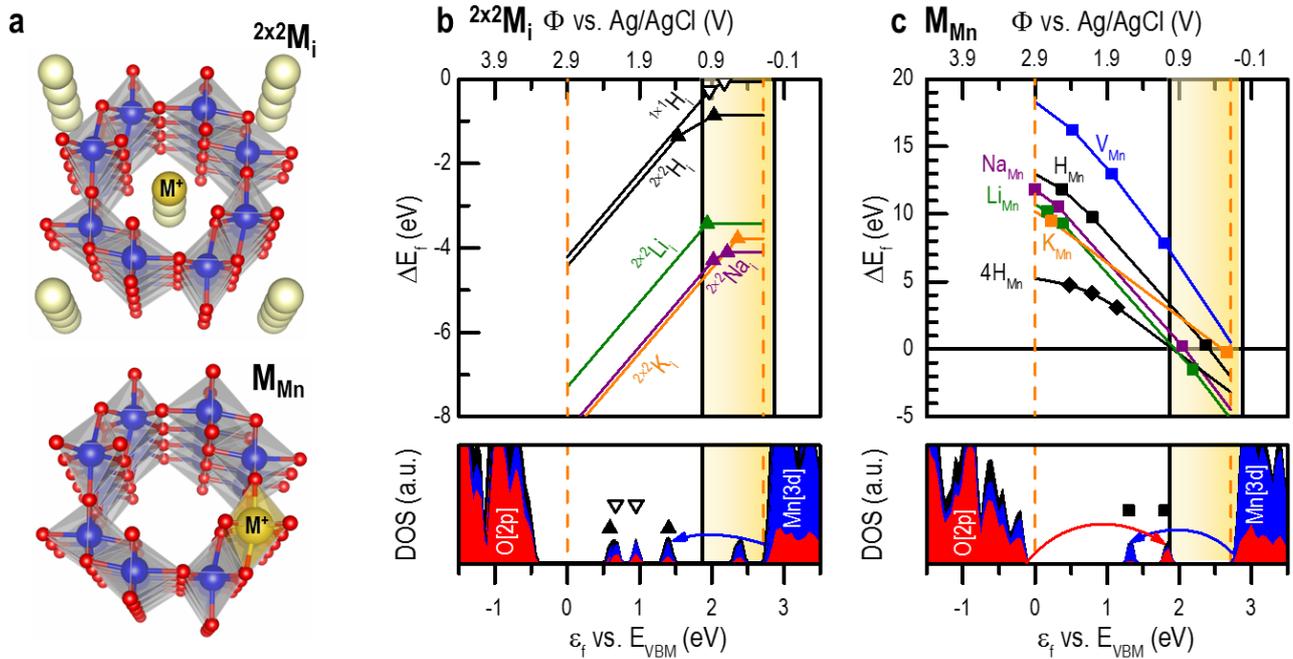

**Figure 4. (a)** Illustration of the mechanism by which interstitial (top) and substitutional (bottom) cations create charge-switching states in α-MnO$_2$. Formation energy and charge-switching state alignments versus potential for cations in **(b)** the 2x2 tunnel and **(c)** Mn vacancies in α-MnO$_2$. The charge-switching states within the band gap are marked for Mn (blue), H (black), Li (green), Na (purple) and K (orange). The states within the scanned potential window consist of interstitial cation defects and cationated Mn vacancies. The dominant orbital character of the states is denoted by the colored arrows in the PDOS shown in the lower panel.

The favorable incorporation of larger cations into the 2×2 tunnels of α-MnO$_2$ is corroborated by the natural existence of cryptomelane (KMn$_8$O$_{16}$) and the spontaneous incorporation of Na$^+$ into α-MnO$_2$ as observed experimentally.[36] These interstitial cations also all induce charge-switching transitions that occur within the SPW, and should therefore all contribute to charge storage in α-MnO$_2$.

**Secondary Substitutional Cation Mechanism**

A secondary mechanism for charge storage in α-MnO$_2$ results from cation substitution at Mn vacancies, as illustrated in Figure 4a. Removing a Mn atom to form a Mn vacancy eliminates interactions between the Mn and neighboring oxygen atoms and creates non-bonding O[2p] orbitals that are destabilized from the valence band into the band gap. For α-MnO$_2$ at a pH of 7.4, H$^+$, Li$^+$, Na$^+$ and K$^+$ all form cationated Mn vacancies that lead to charge-switching states within the SPW, as shown in Figure 4c. When a cation occupies the Mn vacancy, a substitutional defect results with a more negative formation energy than the corresponding Mn vacancy.

The substitutional cation interacts with and stabilizes the nonbonding O[2p] orbitals that resulted from formation of the vacancy, shifting them towards the valence band. The substitutional cation also interacts with Mn-O antibonding states to stabilize them and shift them from the conduction band into the band gap, as shown in Figure 4c. This effect is analogous to the stabilization of Mn-O antibonding states by interstitial cations in the $^{2\times2}M_i$ sites. The radius of the 6-fold coordinated Mn$^{4+}$ center removed to form the vacancy is ~0.5 Å.[37] Consequently, Li$^+$ with a radius of 0.6 Å occupies this vacancy more favorably over the SPW than the larger cations we studied. For more oxidizing conditions our calculations predict that up to four protons will preferentially occupy Mn vacancies creating Ruetschi-type defects.[34] This suggests that a cation exchange process, if kinetically accessible, may also be active over the SPW for substitutional cations. However, the formation of cationated Mn vacancies requires the removal of a sub-surface Mn atom, and the dissolution of Mn$^{2+}$ within the scanned potential window. This process is expected to require a large kinetic barrier that limits the number of these states and consequently reduces their contribution to charge storage.

**Cation Size Effects**

While our results predict that Li$^+$, Na$^+$, and K$^+$ all incorporate at $^{2\times2}M_i$ sites and result in induced states with similar electronic character, our calculations show that cations larger than Na$^+$ are sterically hindered within the 2×2 tunnels of α-MnO$_2$ (Figure 5a). The analysis that led to this conclusion was performed using the ionic radii of 4-fold coordinated cations[37] with the ionic radius of a proton taken to be zero. The maximum passable ionic radius $r_{max}$ is estimated using the calculated diagonal oxygen-to-oxygen distance in the 2×2 tunnel ($l_{OO}$ = 5.01 Å) and the ionic radius for O$^{2-}$ ($r_O$ = 1.21 Å).[37] $r_{max}$ for a 2×2 tunnel was then found to be $^{2\times2}r_{max} = \frac{1}{2}(l_{OO} - 2r_O) = 1.30$ Å. Figure 5b illustrates the effect of the size of the cation species on the formation energy to incorporate it at an interstitial site. The ionic radius of K$^+$ is larger than $r_{max}$ and so the formation energy of $^{2\times2}K_i$ is greater than the formation energy of $^{2\times2}Na_i$, as Na$^+$ is slightly smaller than $r_{max}$.

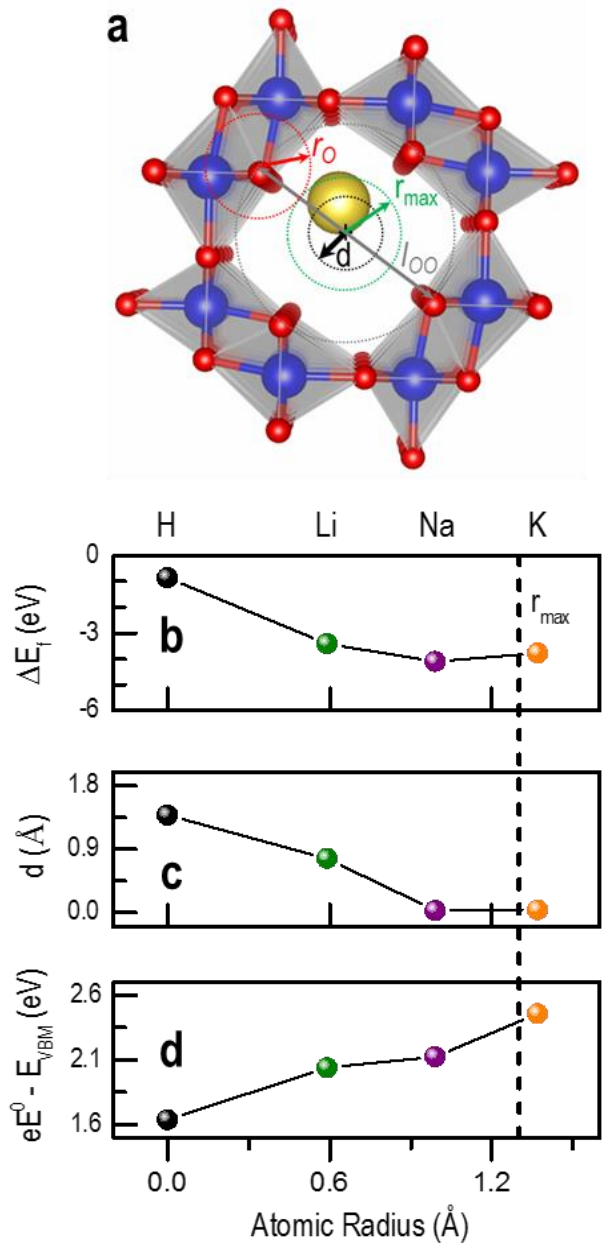

**Figure 5.** Effect of cation size in 2×2 tunnels of α-MnO$_2$. (**a**) Illustration of the parameters used in analyzing cation size effects. (**b**) Effect of the cation species on the formation energy for it to incorporate at an interstitial site. (**c**) Distance between the coordinated cation and the center of the 2×2 tunnel. (**d**) The energy of the most positive charge-switching state for a given interstitial cation plotted relative to the energy of the valence band maximum (VBM).

The distance $d$ between the coordinated cation and the center of the 2×2 tunnel is shown in Figure 5c. Of the 2×2 interstitial cations, H$^+$ and Li$^+$ are located close to the tunnel wall, whereas Na$^+$ and K$^+$ localize near the center of the tunnel and interact more equally with the neighboring oxygen atoms. The charge-switching potential of the most positive charge-switching state for a given interstitial cation is plotted relative to the energy of the valence band maximum in Figure 5d. For example, the Na$^+$ interstitial defect has an electrochemical potential for charge-switching at 2.1 V vs. $E_{VBM}$ or ~0.9V vs Ag/AgCl. As the electrode is scanned past this potential, our calculations show that the interstitial Na$^+$-induced gap states change charge state. All four interstitial cations have different charge-switching potentials resulting from their dissimilar formation energies and their interactions with the neighboring oxygen atoms in the tunnel walls, resulting in capacity differences for different cations as observed experimentally.[17,18,21,38] Na$^+$ has the most favorable interstitial formation energy in the 2×2 tunnels of α-MnO$_2$.

An analogous charge storage mechanism is predicted for the $^{1\times1}M_i$ site where an interstitial cation induces a charge-switching state that undergoes electrochemical reduction to store charge. However, because $^{1\times1}r_{max}$ is only 0.40 Å, protons are the only singly charged cation small enough to occupy $^{1\times1}M_i$ sites. Therefore, only interstitial H$^+$ contributes to charge storage in the 1×1 tunnels. This is corroborated by experimental observations that indicate that both protons and larger cations contribute to charge storage in α-MnO$_2$ where roughly half of the capacity originates from protons.[8,16,18] We calculate that reduction (electron transfer to adjacent Mn-O antibonding orbital) for interstitial $^{1\times1}H^+$ occurs at 0.8 and 0.9 V vs Ag/AgCl. Thus, our calculations predict that as the potential is scanned past these two electrochemical potentials, the proton defect changes charge state by undergoing two discrete one electron reductions, in agreement with XPS results showing the dissociation and formation of hydroxyl groups.[16]

**High Charge Storage Capacity and Rate**

The mechanism of charge storage based on the electrochemical reduction of states induced by interstitial and substitutional cations described above for α-MnO$_2$ at a pH of 7.4 predicts a high charge storage capacity arising from multiple charge-switching states within the SPW and a rapid charge-discharge rate. The interstitial and substitutional mechanisms we identify as leading to charge storage are supported by the large observed capacity of cryptomelane (KMn$_8$O$_{16}$)[39,40] and the enhanced capacity of α-MnO$_2$ when pre-intercalated with large amounts of Na$^+$.[36] In the limit of non-interacting defects and thin-film α-MnO$_2$, the interstitial cation mechanism shown in Figure 4a will result in 1.5 electrons transferred per Mn-center (See SI Section H), with a small additional contribution from the substitutional cation mechanism, also shown in Figure 4a. This accounts for the 1.1 electrons per Mn center observed experimentally for thin-film α-MnO$_2$.[16] For the interstitial cation mechanism, our calculations predict no ion transport through the electrode because protons and larger cations will favorably reside in 1×1 and 2×2 tunnels regardless of the applied potential, as shown in Figure 4b. This computational result of ions favorably residing in the crystal tunnels counters previous assumptions that the charge storage mechanism in α-MnO$_2$ must involve the transport of ions across the double-layer during charge and discharge.[19–21,41] However, our assertion that a fraction of the charge storage capacity exhibited in these materials does not involve ion transport through the bulk electrode or across the double-layer agrees with the high rate of charge-discharge achievable by α-MnO$_2$ electrodes.[19,21,22] Furthermore, the secondary mechanism for charge storage based on substitutional cations (Figure 4a) does involve ion transport across the double-layer as the formation energy for cation incorporation transitions between being favorable and unfavorable over the SPW (Figure 4c). Again, we predict that the majority of the capacity arises from interstitial cations and is thus not mass-transport limited.

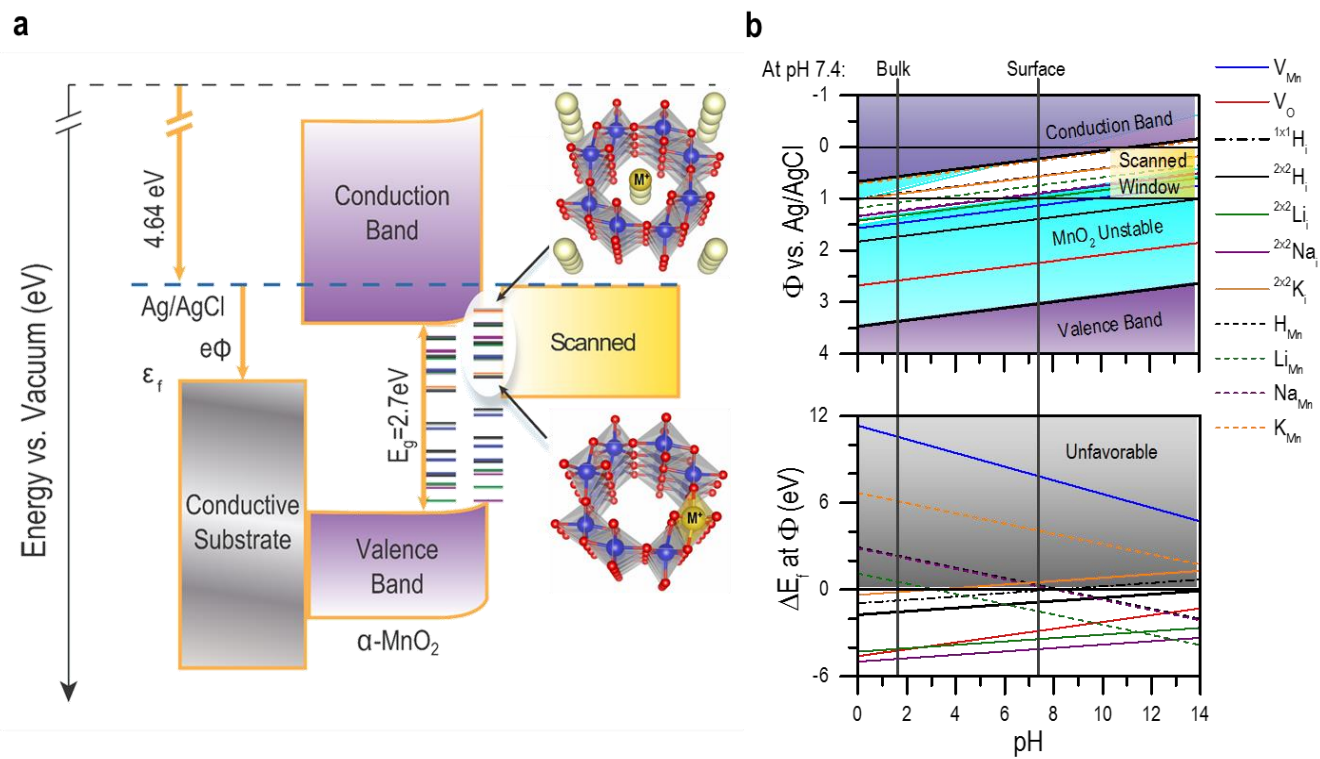

**Figure 6.** Effect of operating conditions on α-MnO$_2$ charge storage. **(a)** Quantitative electrochemical band-diagram description of α-MnO$_2$ in aqueous electrolyte at a pH of 7.4 **(b)** Calculated potential/pH diagram demonstrating the effects of pH on the band alignment and charge-switching potentials of cation incorporation states in α-MnO$_2$ relative to the scanned potential window and the pH dependence of interstitial and substitutional formation energies in α-MnO$_2$.

**Broadening of Peaks Leads to Rectangular CV**

The calculated band diagram and charge-switching states for α-MnO$_2$ are shown in Figure 6a. This analysis follows the framework presented in Figure 1. As an α-MnO$_2$ electrode is scanned from 0 to 1 V versus Ag/AgCl, each charge-switching state appears as a peak on the CV scan. Similarly, the electrons involved in reducing or oxidizing the state appear as current at the applied electrochemical potential in a chronopotentiogram. The charge-switching states that switch charge at potentials within the SPW are the states that contribute to charge storage. However, we expect that the potential at which each charge-switching state undergoes reduction or oxidation shifts depending on the effects of its specific environment. For example, the charge-switching potential will shift as a result of interactions with the solvent and other nearby adsorbed species or defects. Furthermore, the potential at which states change charge will also shift as a function of distance from the surface due to the interfacial field that gives rise to "band bending." These and other effects of the surroundings of the incorporated cations broaden their charge-switching potentials into a distribution spread about each reduction potential, E°. Both the distribution of the equilibrium potentials of the available states and their broadening due to band bending lead to a more rectangular CV — a unique characteristic of pseudocapacitive electrochemical charge storage in α-MnO$_2$.

**Choice of Electrolyte and Operating Conditions**

The effect of electrochemical operating conditions on the charge storage mechanism in α-MnO$_2$ is shown in Figure 6b. The top figure in Figure 6b is similar to a Pourbaix diagram and shows the electrolyte pH (i.e. $\alpha_{H^+}$) sensitivity of the band alignments, charge-switching potentials and stability of MnO$_2$ with respect to the SPW. We find that the operating conditions must be thoroughly considered for optimal charge storage and performance. For α-MnO$_2$, the pH and scanned window that produce the largest capacity in an aqueous electrolyte have been empirically determined, and result in a large capacity contribution from the surface and near-surface where band-bending occurs, and a smaller (but non-zero) capacity contribution from the bulk.

The effect of the electrochemical operating conditions on charge storage in α-MnO$_2$ predicted by our charge storage mechanism explains the higher observed capacitance for thin film α-MnO$_2$ relative to bulk α-MnO$_2$.[16,23] A pH near 7.4 favors MnO$_2$ stability and produces better alignment of the surface and near-surface α-MnO$_2$ band gap that facilitates cycling over a larger fraction of the distribution of charge-switching potentials. In contrast, at a pH of 1.7, the pH$_{PZC}$ value for α-MnO$_2$ (see Methods section), the allowed potential shifts into the equilibrium region for forming Mn$^{2+}$ and has minimal overlap with either the band gap or charge-switching potentials. This potential of zero charge (PZC) pH condition corresponds to the flat band potential of bulk α-MnO$_2$ and indicates that the contribution to capacitance from bulk α-MnO$_2$ is limited relative to thin film α-MnO$_2$ capacities under optimal conditions.

Our calculations predict that the heats of formation are also sensitive to pH such that the concentrations of the different defects can be thermodynamically tuned in consideration of the electrochemical operating conditions (Figure 6b). We stress that the analysis delineated in Figure 6 is applicable to any ion-mediated charge storage material to determine the optimal electrochemical operating conditions that maximize the density of charge-switching states within the SPW while maintaining material stability during cycling. For instance, if a non-aqueous electrolyte with a larger stability window and lower oxidation potential, and an appropriate proton activity were used, the proton-induced charge-switching states of β-MnO$_2$ could be made electrochemically accessible at potentials within the overlap of the band gap and electrolyte window.

## Conclusions

### Insights for Material and Electrolyte Selection

Our model and analysis identifies mechanisms by which a) interstitial cations induce charge-switching states in α-MnO$_2$ through stabilization of Mn-O antibonding orbitals from the α-MnO$_2$ conduction band and b) cations stabilize high energy dangling O[2p] bonds resulting from Mn vacancies. Within the non-interacting defect limit of our approach, these mechanisms will result in 1.5 electrons transferred per Mn-center, which accounts for experimentally observed charge storage of up to 1.1 e$^-$/Mn in aqueous electrolyte. Additionally, the identification of cation-mediated charge storage mechanisms at positive potentials makes α-MnO$_2$ an attractive candidate as a cathode material for lithium ion, sodium ion, and potassium ion batteries.

The charge-switching of these ion-induced band gap states as a mechanism to store charge should be generally applicable to other charge storage materials. For example, other multivalent metal oxides may incorporate interstitial cations that form charge-switching states through stabilization of M-O antibonding orbitals from the conduction band that localize in the band gap. We expect that the multivalent character of the metal center accommodates the localization of charge in these antibonding states leading to their stabilization into the band gap. Additionally, other metal oxides may also contain substitutional cations that interact with high energy O[2p] dangling bonds created by metal center vacancies to form charge-switching states in the band gap. We anticipate that materials could also be developed in which anions shift valence-band states into the band gap by a similar mechanism. For instance, the anion induced charge-switching states created by O$^{2-}$ intercalation may be responsible for the observed charge storage in LaMnO$_3$.[42] If these charge-switching states overlap with the SPW allowed by the electrolyte, those materials will store charge by an analogous mechanism to the one proposed here for α-MnO$_2$.

### Summary

In this work, we have identified an array of physical properties for electrode materials and electrolytes which work in concert to determine the viability of a material to store charge via ion incorporation. These include the band gap, work function, point of zero charge, the tunnel sizes of the electrode material, as well as the pH and stability window of the electrolyte. Our results predict that materials with band gaps that overlap with a given electrolyte window and that have thermodynamically accessible charge-switching states are good candidates as fast and highly reversible charge storage materials. The band edge alignment of many materials with respect to the aqueous electrolyte window have been studied for applications in photoelectrochemical water splitting, and several have been identified as having poor water splitting abilities due to the presence of mid-gap states.[43] We suggest that these poor water splitting materials may be ideal for charge storage if these gap states lie within the SPW of the electrolyte. Computational materials screening applying the framework we have developed here to describe charge storage in α-MnO$_2$ can be employed to rapidly downselect and identify viable new charge storage materials; in line with efforts already in place to computationally identify materials for other electrochemical applications.[44,45] In general, the theoretical model for studying charge storage mechanisms developed and applied in this work should be applicable to the discovery and optimization of new charge storage materials.

## Methods

### Defect Formation Energy Calculations

Electronic structure calculations were performed using density functional theory with a screened non-local exchange-correlation functional (HSE06).[14,15] This functional was calibrated to the results of quasiparticle band structure calculations performed using partially self-consistent GW₀ calculations;[46,47] additional computational details are provided in the SI. Bulk formation energies ($E_f$) for a given defect and charge state were derived from total energy calculations. For a given defect ($D$) in charge state $q$, $E_f$ is expressed in terms of the chemical potentials $\mu_i$ of the defect atom ($i$) and electron ($e$) according to: $\Delta E_f(\mu_i, \mu_e) = E_{tot}^{D,q} - E_{tot}^P - \sum n_i \mu_i + q\mu_e$. Here, $E_{tot}^{D,q}$ and $E_{tot}^P$ are the calculated total energies of the defective ($D$) and perfect ($P$) supercells, respectively, and $n_i$ is the change in the number of atoms of $i$ corresponding to the crystal defect.[10]

The values of $\mu_i$ are constrained by the stability relations of the constituent species necessary for the given phase to exist in equilibrium. For a MnO₂ phase, $\mu_i$ values are bound by Mn-rich and O-rich conditions such that the stability relation $\mu_{Mn} + 2\mu_O = \Delta H_f(MnO_2)$ is satisfied within these bounds. Here, $\Delta H_f(MnO_2)$ is determined from total energy calculations of the perfect crystalline MnO₂ phase of interest and constituent atoms in their reference states. The introduction of H⁺, Li⁺, Na⁺, and K⁺ interstitial or substitutional cation defects were accounted for by including the hydroxide forming limits in the stability relationships $\mu_M + \mu_H + \mu_O = \Delta H_f(MOH)$, where again $\Delta H_f(MOH)$ is determined from total energy calculations of a molecule of MOH and the constituent atoms M, O and H in their reference states. These five stability equations contain six unknown chemical potentials ($\mu_H$, $\mu_{Li}$, $\mu_{Na}$, $\mu_K$, $\mu_O$ and $\mu_{Mn}$). If any unknown $\mu_i$ is specified, the others are determined from the five stability relations. In this work, equilibrium experimental conditions are used to define the $\mu_i$ to predict $\Delta E_f$ relevant to electrochemical supercapacitor (ECSC) operating conditions.

**Chemical Potential of H at the Experimental pH and Applied Bias.**

The value of $\mu_H$ is specified in this work according to the H₂/H⁺ equilibrium relationship under an applied bias as $\mu_H = \frac{1}{2}\mu_{H_2} = \mu_{H^+} + \mu_{e^-}^0 + e\Phi$. Here, $\Phi$ is the applied bias versus a reference, $e$ is a positive elementary charge, $\mu_{e^-}^0$ is the chemical potential of an electron at the reference potential, and $\mu_{H^+}$ is the chemical potential of the proton, which we approximate from the experimental pH by taking $\mu_{H^+} \cong -0.059(pH)$.[48] For aqueous MnO₂ electrochemical supercapacitors, $\Phi$ is commonly scanned from 0.0 to 1.0 V relative to a Ag/AgCl reference electrode. For this work, we reference $\Phi$ to Ag/AgCl by taking the potential of the Standard Hydrogen Electrode (SHE) versus vacuum as 4.44 V[49,50] and Ag/AgCl vs. SHE as 0.197 V,[48] yielding a value of $\mu_{e^-}^0 = -4.637\ eV$. Additionally, the dissolved salt used in aqueous MnO₂ ECSCs is typically a weak base such as KCl or Na₂SO₄ at concentrations ≤ 1 M. Therefore, a slightly basic pH of 7.4 arising from the 0.1 M Na₂SO₄ aqueous electrolyte commonly employed for MnO₂ ECSC systems[23,33] was used to specify $\mu_{H^+} \cong -0.44\ eV$.

Inserting the experimentally defined values of $\mu_{e^-}^0$ and $\mu_{H^+}$ into the H₂/H⁺ equilibrium relationship with an applied bias results in a linear dependence of $\mu_H$ on applied potential ($\mu_H \propto \Phi$). Furthermore, $\Phi$ can be directly related to $\varepsilon_f$, such that $\Delta E_f(\mu_H, \mu_e) = Constant + \varepsilon_f(q+1)$. As a result, the inclusion of an applied potential $\Phi$ into the calculation of $\mu_H$ results in an integer shift in the slope of $\Delta E_f$ vs. $\varepsilon_f$ for each charge state $q$ (SI Section F). This approach differs from previous defect formation studies where $\Delta E_f$ are plotted as a function of $\varepsilon_f$ for a set of fixed $\mu_i$; here $\mu_H$ as a function of $\Phi$ is included in the stability relations to enable calculation of equilibrium defect formation energies $\Delta E_f$ of an electrode material at electrochemical operating conditions.

**Work Functions and Potential of Zero Charge Corrections.**

In order to reference the band gaps of bulk crystalline α-MnO₂ and β-MnO₂ to the SPW, the work functions (W) of these materials must be determined. We calculated W of α- and β- MnO₂ using the electrostatic local potential following the method of Fall *et al*.[51] The change in the local electrostatic potential between the material slab and vacuum ($\Delta V_{el}$) was determined by performing 72-atom unrelaxed slab calculations, while $\varepsilon_f$ was calculated with respect to the average local electrostatic potential using bulk periodic calculations. W of bulk materials depends on the surface termination due to the influence of the electric fields created by surface dipoles of different terminations — for this work, the thermodynamically favorable close-packed (110) surfaces were used to determine W for both α- and β- MnO₂ (SI Section D).[52,53]

When interface conditions produce a net surface charge density, the band edges at the surface shift relative to the PZC, resulting in bending of the electronic bands. A net surface charge produced by acid-base conditions, such as at the experimental pH of 7.4 common in MnO₂ ECSCs, shifts the band edges by an energy determined using the Nernstian relationship $V = V_{PZC} + 0.059(pH_{PZC} - pH)$. We set the value of pH$_{PZC}$ to 7.3 for pyrolusite (natural β-MnO₂) and 1.7 for cryptomelane (α-MnO₂ with potassium doping) to evaluate the band bending corrections due to the electrolyte.[54–56]

## ASSOCIATED CONTENT

### Supporting Information

Supporting Information contains additional computational details of electronic structure calculations, defect formation energies, H chemical potential at an experimental pH with an applied bias, work functions of α- and β- MnO₂, pH derived chemical potentials, theoretical capacity of MnO₂ based on 1 e-transfer per Mn atom and additional diagrams of the defect formation energies of all species at a pH 7.4

## AUTHOR INFORMATION

‡Both authors contributed equivalently

*Corresponding author email: Charles.Musgrave@colorado.edu

## ACKNOWLEDGEMENT

M.J.Y. was supported by a NSF Graduate Research Fellowship under Grant No. DGE 1144083. A.M.H. was supported by NSF CHE-1214131. S.M.G. was supported by DARPA. NSF-MRI Grant CNS-0821794 and the University of Colorado provided computational resources. C.B.M. and S.M.G. are a fellow and affiliate, respectively, of the Renewable and Sustainable Energy Institute and both are fellows of the Materials Science and Engineering Program of the University of Colorado Boulder. We thank Dr. Carl Koval and James Young for useful discussions of electrochemical processes and Erin Breese for designing the graphics of Figures 1 and 6a.

## REFERENCES

(1) Miller, J. R.; Simon, P. *Science* **2008**, *321*, 651–652.
(2) Aricò, A. S.; Bruce, P.; Scrosati, B.; Tarascon, J.-M.; van Schalkwijk, W. *Nat. Mater.* **2005**, *4*, 366–377.
(3) Merlet, C.; Rotenberg, B.; Madden, P. a; Taberna, P.-L.; Simon, P.; Gogotsi, Y.; Salanne, M. *Nat. Mater.* **2012**, *11*, 306–310.
(4) Trasatti, S.; Kurzweil, P. *Platin. Met. Rev.* **1994**, *38*, 46–56.
(5) Song, M.-K.; Cheng, S.; Chen, H.; Qin, W.; Nam, K.-W.; Xu, S.; Yang, X.-Q.; Bongiorno, A.; Lee, J.; Bai, J.; Tyson, T. a; Cho, J.; Liu, M. *Nano Lett.* **2012**, *12*, 3483–3490.
(6) Kötz, R.; Carlen, M. *Electrochim. Acta* **2000**, *45*, 2483–2498.
(7) Xu, C.; Kang, F.; Li, B.; Du, H. *J. Mater. Res.* **2010**, *25*, 1421–1432.
(8) Simon, P.; Gogotsi, Y. *Nat. Mater.* **2008**, *7*, 845–854.
(9) Bélanger, D.; Brousse, L.; Long, J. *Electrochem. Soc. Interface* **2008**.
(10) Zhang, S. B.; Northrup, J. E. *Phys. Rev. Lett.* **1991**, *67*, 2339–2342.
(11) Kohan, a.; Ceder, G.; Morgan, D.; Van de Walle, C. *Phys. Rev. B* **2000**, *61*, 15019–15027.
(12) Radin, M. D.; Siegel, D. J. *Energy Environ. Sci.* **2013**, *6*, 2370.
(13) Holder, A. M.; Osborn, K. D.; Lobb, C. J.; Musgrave, C. B. *Phys. Rev. Lett.* **2013**, *111*, 065901.
(14) Heyd, J.; Scuseria, G. E.; Ernzerhof, M. *J. Chem. Phys.* **2003**, *118*, 8207.
(15) Henderson, T. M.; Janesko, B. G.; Scuseria, G. E. *J. Chem. Phys.* **2008**, *128*, 194105.
(16) Toupin, M.; Brousse, T.; Bélanger, D. *Chem. Mater.* **2004**, *16*, 3184–3190.
(17) Lee, H.; Goodenough, J. *J. Solid State Chem.* **1999**, *223*, 220–223.
(18) Wen, S.; Lee, J.; Yeo, I.; Park, J.; Mho, S. *Electrochim. Acta* **2004**, *50*, 849–855.
(19) Pang, S.-C.; Anderson, M. a.; Chapman, T. W. *J. Electrochem. Soc.* **2000**, *147*, 444.
(20) Chin, S.-F.; Pang, S.-C.; Anderson, M. a. *J. Electrochem. Soc.* **2002**, *149*, A379.
(21) Kuo, S.-L.; Wu, N.-L. *J. Electrochem. Soc.* **2006**, *153*, A1317.
(22) Lee, S. W.; Kim, J.; Chen, S.; Hammond, P. T.; Shao-Horn, Y. *ACS Nano* **2010**, *4*, 3889–3896.
(23) Devaraj, S.; Munichandraiah, N. *J. Phys. Chem. C* **2008**, *112*, 4406–4417.
(24) Brousse, T.; Toupin, M.; Dugas, R.; Athouël, L.; Crosnier, O.; Bélanger, D. *J. Electrochem. Soc.* **2006**, *153*, A2171.
(25) Tompsett, D.; Islam, M. *Chem. Mater.* **2013**.
(26) Tompsett, D.; Parker, S. *Chem. Mater.* **2013**.
(27) Norby, T. *MRS Bull.* **2009**.
(28) Khalil, M. S.; Stoutimore, M. J. a.; Gladchenko, S.; Holder, a. M.; Musgrave, C. B.; Kozen, a. C.; Rubloff, G.; Liu, Y. Q.; Gordon, R. G.; Yum, J. H.; Banerjee, S. K.; Lobb, C. J.; Osborn, K. D. *Appl. Phys. Lett.* **2013**, *103*, 162601.
(29) Ataherian, F.; Wu, N.-L. *J. Electrochem. Soc.* **2011**, *158*, A422.
(30) Ataherian, F.; Lee, K.-T.; Wu, N.-L. *Electrochim. Acta* **2010**, *55*, 7429–7435.
(31) Meng, Y. S.; Arroyo-de Dompablo, M. E. *Energy Environ. Sci.* **2009**, *2*, 589.
(32) Ceder, G.; Supervisor, T.; Schuh, C. **2011**.
(33) Ragupathy, P.; Vasan, H. N.; Munichandraiah, N. *J. Electrochem. Soc.* **2008**, *155*, A34.
(34) Ruetschi, P. *J. Electrochem. Soc.* **1984**, *131*, 2737.
(35) Sun, C.; Zhang, Y.; Song, S.; Xue, D. *J. Appl. Crystallogr.* **2013**, *46*, 1128–1135.
(36) Mai, L.; Li, H.; Zhao, Y.; Xu, L.; Xu, X.; Luo, Y.; Zhang, Z.; Ke, W.; Niu, C.; Zhang, Q. *Sci. Rep.* **2013**, *3*, 1718.
(37) Haynes, W.; Lide, D.; Bruno, T. *CRC Handbook of Chemistry and Physics 2012-2013*; Haynes, W.; Lide, D.; Bruno, T., Eds.; 93rd ed.; CRC Press: New York, 2012.
(38) Ji, C.-C.; Xu, M.-W.; Bao, S.-J.; Cai, C.-J.; Wang, R.-Y.; Jia, D.-Z. *J. Solid State Electrochem.* **2013**, *17*, 1357–1368.
(39) Toupin, M.; Brousse, T.; Bélanger, D. *Chem. Mater.* **2002**, *14*, 3946–3952.
(40) Boisset, A.; Athouël, L. *J. Phys. Chem. C* **2013**, *2*.
(41) Conway, B. E. *Electrochemical supercapacitors: scientific fundamentals and technological applications*; Springer, 1999.
(42) Me, J. T.; Hardin, W. G.; Dai, S.; Johnston, K. P.; Stevenson, K. J. **2014**, *13*.
(43) Bak, T.; Nowotny, J.; Rekas, M.; Sorrell, C. C. *Int. J. Hydrogen Energy* **2010**, *27*, 991–1022.
(44) Greeley, J.; Jaramillo, T. F.; Bonde, J.; Chorkendorff, I. B.; Nørskov, J. K. *Nat. Mater.* **2006**, *5*, 909–913.
(45) Jain, A.; Ong, S. P.; Hautier, G.; Chen, W.; Richards, W. D.; Dacek, S.; Cholia, S.; Gunter, D.; Skinner, D.; Ceder, G.; Persson, K. a. *APL Mater.* **2013**, *1*, 011002.
(46) Shishkin, M.; Kresse, G. *Phys. Rev. B* **2006**, *74*, 035101.
(47) Shishkin, M.; Kresse, G. *Phys. Rev. B* **2007**, *75*, 1–9.
(48) Bard, A. J.; Faulkner, L. R. *Electrochemical Methods: Fundamentals and Applications*; 2nd ed.; Wiley, 2000; p. 856.
(49) Isse, A. a; Gennaro, A. *J. Phys. Chem. B* **2010**, *114*, 7894–7899.
(50) Tripkovic, V.; Björketun, M.; Skúlason, E.; Rossmeisl, J. *Phys. Rev. B* **2011**, *84*, 1–11.
(51) Fall, C.; Binggeli, N.; Baldereschi, A. *J. Phys. Condens. Matter* **1999**, 2689.
(52) Eastment, R.; Mee, C. *J. Phys. F Met. Phys.* **1973**, 1738.
(53) Su, H.-Y.; Gorlin, Y.; Man, I. C.; Calle-Vallejo, F.; Nørskov, J. K.; Jaramillo, T. F.; Rossmeisl, J. *Phys. Chem. Chem. Phys.* **2012**, *14*, 14010–14022.
(54) Sherman, D. M. *Geochim. Cosmochim. Acta* **2005**, *69*, 3249–3255.
(55) Prélot, B.; Poinsignon, C.; Thomas, F.; Schouller, E.; Villiéras, F. *J. Colloid Interface Sci.* **2003**, *257*, 77–84.
(56) Tan, W.; Lu, S.; Liu, F.; Feng, X.; He, J.; Koopal, L. K. *Soil Sci.* **2008**, *173*, 277–286.